\newcommand {\mbf}[1]{\mbox{\boldmath $ #1 $}}
\documentstyle[epsf,12pt]{article}
\newlength{\extraspace}
\newlength{\extraspaces}
\catcode`\@=11
\def\numberbysection{\@addtoreset{equation}{section}
\def\theequation{\arabic{section}.\arabic{equation}}}

\begin{document}
\addtolength{\baselineskip}{.7mm}
\thispagestyle{empty}
\begin{flushright}
{\tt hep-ph/9805395} \\
May, 1998
\end{flushright}
\vspace{2mm}
\begin{center}
{\large{\bf   Finiteness of Multi-Body Neutrino Exchange Potential Energy
}}
\vspace{5pt}
{\large{\bf    in Neutron Stars   }} \\[15mm]
{\sc Jiro Arafune}\footnote{
E-mail: arafune@icrr.u-tokyo.ac.jp}   
 and  
{\sc Yukihiro Mimura}\footnote{
E-mail: mimura@icrr.u-tokyo.ac.jp ;
New address: Theory Group, KEK, Oho 1-1, 
Tsukuba, Ibaraki, 305-0801, Japan}  \\[2mm]
\hspace{2mm}
{\it Institute of Cosmic Ray Research, University of Tokyo \\
Midori-Cho, Tanashi, Tokyo 188-0002, Japan} \\[20mm]

{\bf Abstract}\\[5mm]
{\parbox{14cm}{\hspace{5mm}
The multi-body neutrino potential energy is analytically estimated for a
spherical neutron star  with a vector potential model for neutrinos.  We  
show that the self-energy and the neutrino number of the neutron star coincide  
with the semi-classical values in the large volume limit, and confirm that 
there is no catastrophe in neutron stars with massless neutrinos.
 }}
\end{center}
\vfill
\newpage
\setcounter{section}{0}
\setcounter{equation}{0}
\setcounter{footnote}{0}
\def\theequation{\arabic{section}.\arabic{equation}}
%
%
\vspace{7mm}
\pagebreak[3]
\addtocounter{section}{1}
\setcounter{equation}{0}
\setcounter{subsection}{0}
\setcounter{footnote}{0}
\begin{center}
{\large {\bf \thesection. Introduction}}
\end{center}
\nopagebreak
\medskip
\nopagebreak
\hspace{3mm}
The massless neutrino exchange leads to
a long-range force~\cite{Feinberg}
and in particular to a multi-body potential force
whose astrophysical effects were discussed long ago.~\cite{Hartle,Feynman}
Recently Fischbach \cite{Fisch} argued that the multi-body potential gives  
an unphysically large self-energy to stellar objects like neutron stars and  
concluded that neutrinos should have non-zero masses to resolve this  
paradox.
Soon after that work 
several authors \cite{SV,AGP,KT} made claims 
in contradiction to this conclusion.
Smirnov and Vissani \cite{SV} argued that
the mechanism by which the neutrino sea in neutron stars leading to
blocking of long-range forces
due to the Pauli principle should resolve the paradox.
Abada et al. \cite{AGP} argued that such self-energy is small, exactly  
solving a (1+1) dimensional potential model and a (3+1) dimensional flat  
boarder model.
Kiers and Tytgat \cite{KT}
made numerical analysis of the self-energy and the neutrino number of the  
ground state of a spherical neutron star, and found they tend to
the semi-classical values

\begin{equation}
W_{cl} = - \frac{\mu^4 R^3}{18 \pi},
\label{Wcl}
\end{equation}
\begin{equation}
q_{cl} = \frac{2 (\mu R)^3}{9 \pi}
\label{qcl}
\end{equation}
in the large volume limit and hence that no paradox exists.


 In this paper we calculate the self-energy
and the neutrino number of a spherical neutron star
without recourse to numerical analysis and derive the same conclusion as Kiers  
et al.  based on  a large volume  approximation.
In Section 2, we introduce 
the Schwinger formula for the Weyl spinor and explain our
approximation to calculate
the self-energy and the neutrino number
in the neutron star. In Sectin 3, we give the results of our
calculation.

%
\vspace{7mm}
\pagebreak[3]
\addtocounter{section}{1}
\setcounter{equation}{0}
\setcounter{subsection}{0}
\setcounter{footnote}{0}
\begin{center}
{\large {\bf \thesection. Formulation of Energy and Neutrino Number
}}
\end{center}
\nopagebreak
\medskip
\nopagebreak
\hspace{3mm}
Let us first introduce the Schwinger formula 
for the ground state energy, $W$,  
in the case of a two-component (Weyl) spinor system. 
This is given by the difference  
between the ground state neutrino energy 
of the ``vacuum" containing a neutron star,
$|\hat 0 \rangle$, and that of the true vacuum $| 0 \rangle$. In the  
former, the neutrino propagates in the vector potential inside the neutron star  
due to $Z^0$ exchange,
\begin{equation}
W = \langle \hat 0 | H | \hat 0 \rangle -
\langle 0 | H_0 | 0 \rangle ,
\label{W1}
\end{equation}
where $H_0$ is the free Hamiltonian, and $H$ is the total Hamiltonian for the  
neutrinos with the neutron star. The latter corresponds to the Lagrangian
for the (2-component)
neutrino field $\chi$ and its hermitian conjugate $\bar\chi$,
\begin{equation}
{\cal L} = \bar\chi i \sigma^\mu \partial_\mu \chi +
\frac{G_W}{\sqrt 2} \bar\chi \sigma^\mu \chi j_\mu,
\end{equation}
with $\sigma^0 =\sigma_0=1$, and
$\sigma^i=-\sigma_i$ ($i=1,2,3$)  the Pauli matrices.
Here $j_\mu$ is the weak current of the neutrons proportional to the neutron  
density,
\begin{equation}
j_\mu \sim \langle n^\dagger n \rangle g_{\mu 0}.
\end{equation}
In the following we consider the model Lagrangian of the
neutron stars \cite{AGP}
\begin{equation}
{\cal L} = \bar\chi ( i \sigma^\mu \partial_\mu + \mu(x) ) \chi,
\end{equation}
where
\begin{equation}
\mu(x) = \mu\,\theta(R - |{\mbf x}|),
\end{equation}
with $R$  the neutron star radius and  $ \mu$ =
$\frac{G_W}{\sqrt 2}  \langle n^\dagger n \rangle $ a constant, typically
on the order of several eV.

Since the Hamiltonian densities in Eq.~(\ref{W1}) are given by  
$\displaystyle
-i\lim_{x'\rightarrow x}\frac{\partial}{\partial x_0} {\rm \, tr} {\chi  
(x)\bar\chi(x')}$,
the Schwinger formula for the energy is given by
\begin{eqnarray}
W &=& - i \int d^3 x \frac{\partial}{\partial x_0}
{\rm \, tr} (S(x,x')- S^0(x,x'))_{x'\rightarrow x}
\label{W2}\\
&=& - \frac{1}{2\pi i} \int dE  {\rm \, Tr} (\ln (E-H) - \ln (E-H_0)),
\label{Schform}
\end{eqnarray}
where the symbol Tr is the trace over both spinor and configuration
space indices,
$S(x,x')$ is the Feynman propagator
defined as
\begin{equation}
(i \sigma \partial + \mu(x)) S(x,x') = i \delta(x,x'),
\end{equation}
\begin{equation}
\langle 0| T \chi(x) \bar\chi(x')|0 \rangle = S(x,x'),
\end{equation}
and $S^0(x,x')$ is that of the free neutrino.
In Eq.~(\ref{Schform}) and thereafter 
we use the same notation $H$ for the quantum mechanical Hamiltonian,
$H = -i{\mbf \sigma \nabla}-\mu (x)$, 
as for the field theoretical one in Eq.~(\ref{W1}), 
since this should have no confusion. 
To derive Eq.~(\ref{Schform}) from Eq.~(\ref{W2}) we have used the formula
\begin{equation}
S(x,x')= - \frac{1}{2\pi i} \lim_{\epsilon\rightarrow +0}{\int dE  \langle x|   
\frac{1}{E(1+i \epsilon)-H}| x'\rangle e^{-iE(x^0-x'^0)}},
\label{S}
\end{equation}
and a similar one for $S^0(x,x')$.  
The prescription with $i \epsilon$ is introduced  
here to satisfy the boundary condition of the time ordering propagator.
The neutrino number, $q$,  is similarly given as
\begin{eqnarray}
q &=& - \int d^3 x  {\rm \, tr} (S(x,x')- S^0(x,x'))_{x'\rightarrow x} \\
&=& \frac1{2\pi i} \int dE {\rm \, Tr} \left(\frac1{E-H} - \frac1{E-H_0}\right).
\label{nu-number}
\end{eqnarray}

In order to avoid the apparent divergence of the expressions for $W$  
and $q$, we differentiate the energy with respect to $\mu$ twice and the  
neutrino number once to obtain
\begin{equation}
\frac{d^2W}{d\mu^2} =
\frac1{2\pi i} \int dE \int d^3 x \int d^3 y
\ {\rm tr} \left(\frac1{E-H}\right)_{{\mbf x},{\mbf y}} \theta(R-|{\mbf x}|)
\left(\frac1{E-H}\right)_{{\mbf y},{\mbf x}} \theta(R-|{\mbf y}|),
\label{d2W/dmu2}
\end{equation}
\begin{equation}
\frac{d q}{d\mu} =
- \frac{1}{2\pi i} \int dE
\int d^3 x \int d^3 y \left(\frac1{E-H}\right)_{{\mbf x},{\mbf y}}
\left(\frac1{E-H}\right)_{{\mbf y},{\mbf x}} \theta(R-|{\mbf x}|).
\label{dq/dmu}
\end{equation}

%
%
In Eq.~(\ref{d2W/dmu2}) we see that the neutrino propagates from a point $\mbf  
x$ inside the  neutron star to another point $\mbf y$ also inside the  
neutron star.   Noting  $\mu R \sim 10^{12}$, we see that the neutrino  
propagator rapidly oscillates in the neutron star, and thus only a 
short distance  
with  $|{\mbf x} - {\mbf y}| \ll  R$ or finite $\mu |{\mbf x} - {\mbf y}|$ contributes to  
the integration of Eq.~(\ref{d2W/dmu2}) and (\ref{dq/dmu}).  Thus it is
safe to replace  the full propagator of the neutrino
\begin{equation}
\left(\frac1{E-H_0+\mu\, \theta(R-r)}\right)_{{\mbf x},{\mbf y}}
\end{equation}
with the approximate one
\begin{equation}
\left(\frac1{E-H_0+\mu}\right)_{{\mbf x},{\mbf y}}
\end{equation}
to evaluate Eqs.~(\ref{d2W/dmu2}) or (\ref{dq/dmu}), neglecting
the small contribution of such neutrinos that virtually propagate out of the  
neutron star (see Fig.~1).

\begin{figure}[htbp]
 \leavevmode
 \epsfysize=3cm
\centerline{\epsfbox{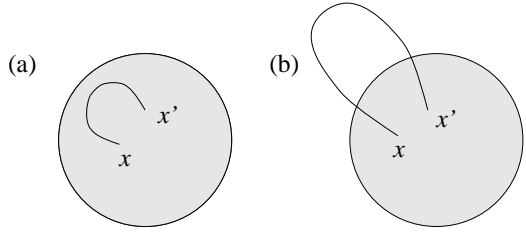}}

 \caption{We neglect the neutrino propagating through
          path (b).
          We only take the path (a) into account.
}
 \label{fig:1}
\end{figure}

This approximation\footnote{
Here the $i \epsilon$ prescription is similar to that used in Eq.~(\ref{S}). }
leads to
\begin{eqnarray}
\left(\frac1{E-H_0+\mu\, \theta(R-r)}\right)_{{\mbf x},{\mbf y}}
& \approx &  \left(\frac1{E-H_0+\mu}\right)_{{\mbf x},{\mbf y}}  \nonumber\\
&=&  \int \frac{d^3 p}{(2\pi)^3}
\frac1{E+\mu - \mbf{\sigma p}} e^{i \mbf{p} ({\mbf x}-{\mbf y})} \nonumber \\
&=&  -\left(E+\mu +  \frac{\mbf{\sigma r}}{r} 
\left( \frac{i}{r} + p^+\right)\right)
\frac{e^{i p^+ r}}{4\pi r},
\label{propa}
\end{eqnarray}
where ${\mbf r} = {\mbf x} - {\mbf y}$ and
$p^+ = \varepsilon(E) (E+\mu)$ ($\varepsilon(E) = 1$ $(-1)$ for
$E>0$ ($E<0$)).
The Eq.~(\ref{propa}) represents 
nothing but a full propagator in condensed matter  
of infinite volume.

The Schwinger formula (\ref{Schform}) has an ultraviolet divergence  
proportional to $\mu^2$, which corresponds to just the familiar vacuum
polarization diagram for $Z^0$  \cite{KT} 
and should be subtracted.\footnote{
We note that a divergence cannot be subtracted in case of an  
electron-condensate medium,
since the exchange of charged weak boson is involved. 
In such a case we need a  
cut-off scale $\Lambda_c$,
which is the  weak boson mass or the inverse of the mean distance of the  
electrons.}
Thus, the right-hand side of Eq.~(\ref{d2W/dmu2}), 
a second derivative of $W$, has a  
constant divergence which should be subtracted.
Equation (\ref{nu-number}) for the neutrino number
also has a divergence. This divergence, linear in $\mu$, corresponds to a  
one-loop correction to the neutrino density operator, familiar in the  
correction in the operator product expansion, and is caused by the operator  
mixing of the neutrino density and neutron density. This is also to be  
subtracted by renormalization. Thus the right-hand side 
of Eq.~(\ref{dq/dmu}), the first  
derivative of $q$,  should be renormalized to be zero at $\mu = 0$.

\newpage
%
%
\vspace{7mm}
\pagebreak[3]
\addtocounter{section}{1}
\setcounter{equation}{0}
\setcounter{subsection}{0}
\setcounter{footnote}{0}
\begin{center}
{\large {\bf \thesection.  Evaluation of Energy and Neutrino Number
}}
\end{center}
\nopagebreak
\medskip
\nopagebreak
\hspace{3mm}
In this section, we evaluate the self-energy and neutrino number of a  
neutron star with Eqs.~(\ref{d2W/dmu2}) and (\ref{dq/dmu}) using the  
approximation Eq.~(\ref{propa}). It is easy to formally obtain
\begin{equation}
\frac{d^2W}{d\mu^2}
=\frac 1{16{\pi}^3 } 
{\int}_{|{\mbf x}| \leq R} d^3 x {\int}_{|{\mbf y}| \leq R} d^3  y
\ r^{-5}(\cos 2\mu r+\mu r \sin 2\mu r),
\end{equation}
with ${\mbf r}={\mbf y} -{\mbf x}$.
We renormalize it by subtracting the divergent term to obtain
\begin{eqnarray}
\left.\frac{d^2W}{d\mu^2} \right|_{\rm {ren}}
&=& \frac{d^2W}{d\mu^2} - \left.\frac{d^2W}{d\mu^2} \right|_{\mu =0} \\
&=& \frac 1{8{\pi}^3 }{\int}_{|{\mbf x}| \leq R} d^3 x {\int}_{|{\mbf y}| \leq R}  
d^3 y \  r^{-5}(\cos 2\mu r+\mu r \sin 2\mu r -1).
\label{RenW1}
\end{eqnarray}
Noting the convergence of the integrand of Eq.~(\ref{RenW1}) 
both at $r=0$ and  
$r=\infty$,  we see that the change of the integration region
${\int}_{|{\mbf x}| \leq R} d^3 x {\int}_{|{\mbf y}| \leq R} d^3 r$  $\rightarrow$
${\int}_{|{\mbf x}| \leq R} d^3 x {\int}_{|{\mbf r}| < \infty} d^3 r$
is allowed in the large volume limit. 
Then the integration in Eq.~(\ref{RenW1})  
is analytically performed to give
\begin{eqnarray}
\left. \frac{d^2W}{d\mu^2} \right|_{\rm {ren}}
& \approx & \frac 1{8{\pi } ^3}{\int}_{|{\mbf x}| \leq R} d^3 x {\int}_{r <  
\infty} 4\pi r^2 dr \ r^{-5}(\cos 2\mu r+\mu r \sin 2\mu r -1)
\label{RenW1b}\\
&= & \frac 1{8{\pi}^3 } {\int}_{|{\mbf x}| \leq R} d^3 x (-4\pi {\mu}^2)\\
&=& - \frac 2{3\pi }{\mu}^2R^3.
\label{RenW2}
\end{eqnarray}
By integration we obtain the renormalized self-energy
\begin{equation}
W  = -\frac 1{18\pi }{\mu}^4R^3,
\label{RenW3}
\end{equation}
where we have used the fact \cite{KT}  
that $W$ is an even function of $\mu $ and  
$\displaystyle W|_{\mu =0}= \left.\frac{d^2W}{d\mu^2}\right|_{\mu =0} =0$. The result  
Eq.~(\ref{RenW3}) coincides with the semi-classical value of Eq.~(\ref{Wcl}).

Here we make some remarks regarding some subtleties of the calculation. 
If we differentiate $W$ with  
respect to $\mu$ more times, we obtain better convergence at short  
distance.  In this sense it is safer to evaluate $W$ in terms of higher  
derivatives. In fact we have no divergence in $W^{(n)}(\mu)$ for $n \geq 3$.  
In the case of  $W^{(1)}(\mu)$, apparently we may have divergence worse than in  
the case of $W^{(2)}$. When we calculate
$W^{(1)}(\mu)$ and use the approximation Eq.~(\ref{propa}), 
however, we obtain a finite result,
\begin{eqnarray}
\frac{dW}{d\mu} &=& \frac1{2\pi i} \int^\infty_{-\infty} dE
\int_{|x|<R} d^3 x
\ 2 (E+\mu) \frac{ e^{ip^+ r}}{4\pi r} |_{r \rightarrow 0} \\
&\approx & \frac1{2 \pi^2} \int_{|x|<R} d^3 x \ 
\frac{\mu r \cos \mu r - \sin \mu r}{r^3} |_{r \rightarrow 0}
\label{superficial} \\
&=& -\frac2{9\pi} \mu^3 R^3,
\end{eqnarray}
which is consistent with the renormalized one, Eq.~(\ref{RenW3}) .
The ultraviolet divergence linear in  $\mu$
is accidentally cancelled in this calculation.
Since Eq.~(\ref{superficial}) has a superficially divergent dimension,
it requires a more careful calculation by taking account of the boarder  
effect.
In order to study the boarder effect, let us take $\displaystyle  
\frac{d^2W}{d\mu dR}$,
\begin{eqnarray}
\frac{d^2 W}{d\mu dR} &=&
-\frac1{2\pi i} \int dE \ {\rm Tr} 
\left(\frac1{E-H} \ \delta(R-|{\mbf x}|)\right) \nonumber
\\
&+ &\ \frac{\mu}{2\pi i} \int dE \ {\rm tr} \int d^3 x \int d^3 y
\left(\frac1{E-H}\right)_{{\mbf x},{\mbf y}}\! \delta(R-|{\mbf y}|)
\left(\frac1{E-H}\right)_{{\mbf y},{\mbf x}}\! \theta(R-|{\mbf x}|). \nonumber
\label{d2W/dmudR} \\
\end{eqnarray}
We note this expression is exact.
We again approximate it  with the full propagator
by Eq.~(\ref{propa}). The result has an ultraviolet divergence linear in   
$\mu$.

After renormalizing as explained above, we obtain
\begin{equation}
\frac{d^2 W}{d\mu dR} \mid_{\rm ren} =
-\frac{2}{3\pi} \mu^3 R^2 +
\mu \,\frac{\cos 4\mu R + 4 \mu R \ {\rm Si}(4\mu R)}{8 \pi R},
\end{equation}
where $\displaystyle {\rm Si}(x) = \int^x_0 dt \frac{\sin t}{t}$. This gives
 the self-energy of the neutron star as
\begin{equation}
W = -\frac{\mu^4 R^3}{18 \pi} \left( 1 + O\left(\frac1{R}\right)\right),
\end{equation}
which again coincides with the classical value Eq.~(\ref{Wcl}) 
in the large $R$ limit.

Now let us evaluate the neutrino number. Equation (\ref{dq/dmu}) for  
$\displaystyle \frac{d q}{d\mu}$ 
resembles Eq.~(\ref{d2W/dmu2}) for $\displaystyle \frac{d^2W}{d\mu^2} $,
except for the sign and the range of
the $y$ integration.
Since the quantity
\begin{equation}
\int_{|{\mbf x}|< R} d^3 x \int_{|{\mbf y}|>R} d^3 y
{\rm \, tr} \left(\frac1{E-H}\right)_{{\mbf x},{\mbf y}}
\left(\frac1{E-H}\right)_{{\mbf y},{\mbf x}}
\end{equation}
should be negligibly small for large $\mu R$,
they should coincide (except for the sign) to give
\begin{equation}
\left.\frac{dq}{d\mu} \right|_{\rm ren}
= - \left.\frac{d^2W}{d{\mu}^2} \right|_{\rm ren} 
=  \frac{2}{3\pi} \mu^2 R^3,
\end{equation}
with the aid of Eq.~(\ref{RenW2}). This gives the neutrino number,
\begin{equation}
q = \frac{2}{9\pi} \mu^3 R^3,
\end{equation}
which is equal to the semi-classical value in Eq.~(\ref{qcl}).

%
%
\vspace{7mm}
\pagebreak[3]
\addtocounter{section}{1}
\setcounter{equation}{0}
\setcounter{subsection}{0}
\setcounter{footnote}{0}
\begin{center}
{\large {\bf \thesection. Discussion
}}
\end{center}
\nopagebreak
\medskip
\nopagebreak
\hspace{3mm}
In this paper we estimated the self energy
and the neutrino number of  neutron stars
without recourse to numerical analysis, and confirmed
that they coincide with the semi-classical values in the large volume limit.  
They do not lead to paradox nor to any lower bound for the neutrino masses.  
While our estimation was made non-perturbatively,  it  may be instructive to  
evaluate the self-energy  perturbatively by expanding the integrand of  
Eq.~(\ref{RenW1}) with respect to $\mu$ and integrating each term over $\mbf x$  
and $\mbf y$ without using the approximation Eq.~(\ref{RenW1b}).
The integration is easy to perform and  
leads to  an alternating series in powers  
of $\mu R$. This series is similar (though not  exactly equal) to that  
considered in Ref.~\cite{Fisch}
and an example where only the sum (but not each  
term individually) is meaningful in large $R$ limit.

%

\begin{center}
{\large {\bf  Note Added in Proof
}}
\end{center}

In Ref.~\cite{AGP} Abada et al. estimated the self energy
and the neutrino number by independent methods and got results 
different from ours. 
Their self energy is linear in the external potential ($\mu$ in our notation), 
while it should be even to keep CP invariance. 
We think that the difference is due to their incomplete Hamiltonian 
violating CP invariance without taking the average of it and its CP conjugation.
Our method keeps the invariance, though implicitly, 
by taking the neutrino propagator with adequate $i \epsilon$ procedure. 
Kiers et al. \cite{KT} also analyzed them independently 
with the result in good agreement with ours.

%
\pagebreak[3]
\addtocounter{section}{1}
\setcounter{equation}{0}
\setcounter{subsection}{0}
\setcounter{footnote}{0}
\begin{center}
{\large {\bf  Acknowledgement
}}
\end{center}
\nopagebreak
\medskip
\nopagebreak
\hspace{3mm}
The authors are grateful to V. Rubakov for valuable comments and to
E. Fischbach,  
M. Koike, and K. Ogure  for stimulating discussions.
\vspace{5mm}
%
\newcommand{\NP}[1]{{\it Nucl.\ Phys.\ }{\bf #1}}
\newcommand{\PL}[1]{{\it Phys.\ Lett.\ }{\bf #1}}
\newcommand{\CMP}[1]{{\it Commun.\ Math.\ Phys.\ }{\bf #1}}
\newcommand{\MPL}[1]{{\it Mod.\ Phys.\ Lett.\ }{\bf #1}}
\newcommand{\IJMP}[1]{{\it Int.\ J. Mod.\ Phys.\ }{\bf #1}}
\newcommand{\PRP}[1]{{\it Phys.\ Rep.\ }{\bf #1}}
\newcommand{\PR}[1]{{\it Phys.\ Rev.\ }{\bf #1}}
\newcommand{\PRL}[1]{{\it Phys.\ Rev.\ Lett.\ }{\bf #1}}
\newcommand{\PTP}[1]{{\it Prog.\ Theor.\ Phys.\ }{\bf #1}}
\newcommand{\PTPS}[1]{{\it Prog.\ Theor.\ Phys.\ Suppl.\ }{\bf #1}}
\newcommand{\AP}[1]{{\it Ann.\ Phys.\ }{\bf #1}}
\newcommand{\ZP}[1]{{\it Zeit.\ f.\ Phys.\ }{\bf #1}}


\begin{thebibliography}{100}
%
\bibitem{Feinberg} G. Feinberg and J. Sucher,  \PR{166} (1968) 1638.
%
\bibitem{Hartle} J.B. Hartle, \PR{D1} (1970) 394.
%
\bibitem{Feynman} R. Feynman,  in  {\it Feynman Lectures on Gravitaion,} 
Ed. R. Feynman et al. (Addison-Wesley, 1995, 232 p.).
%
\bibitem{Fisch} E. Fischbach, \AP{247} (1996) 213.
%
\bibitem{SV} A.Y. Smirnov and F. Vissani, hep-ph/9604443.
%
\bibitem{AGP} A. Abada, M. B. Gavela and O. P\`ene,
\PL{B387} (1996) 315. \\
A. Abada, O. P\`ene and J. Rodr\'\i guez-Quintero,
\PL{B423} (1998) 355; \PR{D58} (1998) 073001.
%
\bibitem{KT} K. Kiers and M. H. G. Tytgat, \PR{D57} (1998) 5970;
hep-ph/9807412.
%
\bibitem{Schwin} J. Schwinger, \PR{94} (1954) 1362.
%
\end{thebibliography}
\end{document}